\documentclass[11pt,a4paper]{article}
\pdfoutput=1

\usepackage[nosort]{cite}
\usepackage[bulletsep]{collref}

\usepackage[british]{babel} %dates in british format

\usepackage{amsmath,amssymb,amsthm}
\usepackage{mathrsfs}
\usepackage{mathtools}
\usepackage{bbm}
\usepackage{bm}
\usepackage{slashed}
\usepackage{braket}
\usepackage{dsfont}

\usepackage{color}
\usepackage[textwidth = 430 pt, textheight = 630 pt]{geometry}
\definecolor{MyDarkBlue}{rgb}{0.15,0.25,0.45}

\usepackage[linktocpage=true,hypertexnames=false]{hyperref}
\hypersetup{
colorlinks=true,
citecolor=MyDarkBlue,
linkcolor=MyDarkBlue,
urlcolor=MyDarkBlue,
pdfauthor={Antonio Pittelli, Alessandro Torrielli, and Martin Wolf},
pdftitle={Secret Symmetries of Type IIB Superstring Theory on AdS3 x S3 x M4},
pdfsubject={hep-th}
breaklinks=true
}

\let\fn\footnote
\renewcommand{\footnote}[1]{\linespread{1.1}\fn{#1}\linespread{1.29}}

\makeatletter\renewcommand{\section}{\@startsection
{section}{1}{\z@}{-3.5ex plus -1ex minus
    -.2ex}{2.3ex plus .2ex}{\bf\mathversion{bold} }}
\makeatletter\renewcommand{\subsection}{\@startsection{subsection}{2}{\z@}{-3.25ex
plus -1ex minus
   -.2ex}{1.5ex plus .2ex}{\bf\mathversion{bold} }}
\makeatletter\renewcommand{\subsubsection}{\@startsection{subsubsection}{3}{-2.45ex}{-3.25ex
plus -1ex minus -.2ex}{1.5ex plus .2ex}{\it }}
\renewcommand{\thesection}{\arabic{section}}
\renewcommand{\thesubsection}{\arabic{section}.\arabic{subsection}}
\renewcommand{\@seccntformat}[1]{\@nameuse{the#1}.~~}

\renewcommand{\theequation}{\thesection.\arabic{equation}}
\makeatletter \@addtoreset{equation}{section}

\renewcommand*\l@section{\@dottedtocline{1}{0em}{2em}}
\renewcommand*\l@subsection{\@dottedtocline{2}{2em}{2.4em}}
\renewcommand*\l@subsubsection{\@dottedtocline{4}{3.8em}{3.7em}}

\renewcommand\tableofcontents{%
    \section*{\large\contentsname
        \@mkboth{%
          \MakeUppercase\contentsname}{\MakeUppercase\contentsname}}%
       {\baselineskip=15pt plus 2pt minus 1pt
    \@starttoc{toc}}%
}

\renewenvironment{thebibliography}[1]
     {\baselineskip=16pt plus 2pt minus 1pt
      \section*{\large\refname
        \@mkboth{\MakeUppercase\refname}{\MakeUppercase\refname}}%
     \list{\@biblabel{\@arabic\c@enumiv}}%
           {\settowidth\labelwidth{\@biblabel{#1}}%
            \leftmargin\labelwidth
            \advance\leftmargin\labelsep
            \@openbib@code
            \usecounter{enumiv}%
            \let\p@enumiv\@empty
            \renewcommand\theenumiv{\@arabic\c@enumiv}}%
      \sloppy
      \clubpenalty4000
      \@clubpenalty \clubpenalty
      \widowpenalty4000%
      \sfcode`\.\@m
 \catcode`\^^M=10%
}

\newcommand{\appendices}{
\section*{Appendix}\label{appendices}\setcounter{subsection}{0}
\addcontentsline{toc}{section}{Appendix}
\setcounter{equation}{0}
\makeatletter
\renewcommand{\theequation}{\Alph{subsection}.\arabic{equation}}
\renewcommand{\thesubsection}{\Alph{subsection}}
\@addtoreset{equation}{subsection}
\makeatother
}

\newcommand{\eand}{{~~~\mbox{and}~~~}}     		
     		
\newcommand{\ewith}{{~~~\mbox{with}~~~}}
\newcommand{\efor}{{~~~\mbox{for}~~~}}
\newcommand{\eforall}{{~~~\mbox{for all}~~~}}
\newcommand{\di}{\text{i}}     			
\newcommand{\de}{\text{e}}   
\newcommand{\bA}{{\bf A}}     		
\newcommand{\bB}{{\bf B}}     		
\newcommand{\bC}{{\bf C}}     		
\newcommand{\bD}{{\bf D}}

\newcommand{\bb}[1]{\mathbb{#1}}
\newcommand{\be}[1]{\begin{equation}#1\end{equation}}

\renewcommand{\geq}{\geqslant}

\DeclareMathOperator{\str}{str}

\def\XXint#1#2#3{{\setbox0=\hbox{$#1{#2#3}{\int}$}
    \vcenter{\hbox{$#2#3$}}\kern-.5\wd0}}

\newcommand{\AdS}{\text{AdS}}
\newcommand{\CFT}{\text{CFT}}
\newcommand{\Sphere}{S}
\newcommand{\Torus}{T}

\newcommand{\matId}{\mathds{1}}

\newcommand{\comm}[2]{[#1,#2]}
\newcommand{\acomm}[2]{\{#1,#2\}}

\newcommand{\alg}[1]{\mathfrak{#1}}

\newcommand{\algSU}{\alg{su}}

\newcommand{\algU}{\alg{u}}

\newcommand{\gen}[1]{\mathfrak{#1}}
\newcommand{\Ygen}[1]{#1}

\newcommand{\ccomm}[1]{\left\{#1\right\}}

\newcommand{\gcomm}[1]{\left[#1\right\}}
\newcommand{\smallL}{\scriptscriptstyle\textit{L}}
\newcommand{\smallR}{\scriptscriptstyle\textit{R}}

\newcommand{\smallRR}{\scriptscriptstyle\textit{RR}}

\newcommand{\p}[1]{\left(#1\right)}

\newcommand{\id}{\mathbbm 1}

\newcommand{\superN}{\mathcal{N}}

\begin{document}

\begin{titlepage}

\setcounter{page}{0}
\renewcommand{\thefootnote}{\fnsymbol{footnote}}

\begin{flushright}
DMUS--MP--14/03
\end{flushright}

\vspace{1cm}

\begin{center}

\textbf{\LARGE\mathversion{bold} Secret Symmetries of Type IIB Superstring Theory on $\AdS_3\times S^3\times M^4$}

\vspace{1cm}

{\large Antonio Pittelli, Alessandro Torrielli, and Martin Wolf \footnote{{\it E-mail addresses:\/}
\href{mailto:a.pittelli@surrey.ac.uk}{\ttfamily a.pittelli@surrey.ac.uk},
\href{mailto:a.torrielli@surrey.ac.uk}{\ttfamily a.torrielli@surrey.ac.uk}, 
\href{mailto:m.wolf@surrey.ac.uk}{\ttfamily m.wolf@surrey.ac.uk}}
} 

\vspace{1cm}

{\it Department of Mathematics, University of Surrey\\
Guildford GU2 7XH, United Kingdom}

\vspace{1cm}

{\bf Abstract}
\end{center}
\vspace{-.3cm}
\begin{quote}
We establish features of so-called Yangian secret symmetries for $\AdS_3$ type IIB superstring backgrounds thus verifying the persistence of such symmetries to this new instance of the AdS/CFT correspondence. Specifically, we find two {\it a priori} different classes of secret symmetry generators. One class of generators,  anticipated from the previous literature,  is more naturally embedded in the algebra governing the integrable scattering problem. The other class of generators is more elusive, and somewhat closer in its form to its higher-dimensional $\AdS_5$ counterpart. All of these symmetries respect left-right crossing. In addition, by considering the interplay between left and right representations, we gain a new perspective on the $\AdS_5$ case. We also study the $R\mathcal T\mathcal T$-realisation of the Yangian in $\AdS_3$ backgrounds thus establishing a new incarnation of the Beisert--de Leeuw construction.
\vfill
\noindent 29th October 2014

\end{quote}

\setcounter{footnote}{0}\renewcommand{\thefootnote}{\arabic{thefootnote}}

\end{titlepage}

\tableofcontents

\bigskip
\bigskip
\hrule
\bigskip
\bigskip

\section{Introduction, summary, and outlook}
\label{intro}

\subsection{Introduction}

\paragraph{Secret symmetries.}
The integrable structure permeating the AdS/CFT correspondence (see {\it e.g.}~\cite{Beisert:2010jr,Arutyunov:2009ga} for reviews) keeps revealing new surprising features that extend our algebraic understanding of scattering problems and uncover new structures in supersymmetric quantum groups.    

For instance, the Hopf superalgebra which controls the AdS${}_5$/CFT${}_4$ integrable system is a rather unconventional infinite-dimensional Yangian-type symmetry \cite{Drin,Chari,Khoroshkin:1994uk,MacKay:2004tc,Molev}, whose level-0 Lie algebra sector is Beisert's three-fold centrally-extended $\alg{psu}(2|2)$ superalgebra \cite{Beisert:2005tm,Arutyunov:2006ak}. At level 1, the Yangian generators\footnote{For supersymmetric Yangians, the reader is referred {\it e.g.}~to \cite{Yao,stuko,Gow,Spill:2008yr}.} come each as partners to the level-0 ones, except for the presence of an additional symmetry \cite{Matsumoto:2007rh,Beisert:2007ty}, which has no analog at level 0. This {\it secret} generator corresponds to a hypercharge, and acts as a fermion number on the scattering particles, counting the total number of fermions. Specific entries in Beisert's $S$-matrix \cite{Beisert:2005tm} of the form $|\mbox{boson} \rangle \otimes |\mbox{boson} \rangle \mapsto |\mbox{fermion} \rangle \otimes |\mbox{fermion} \rangle$ and vice versa, for example, break the fermion number $(-)^F \otimes \matId + \matId \otimes (-)^F$; here, $F$ denotes the fermion number operator. This is restored at level 1 by means of the non-trivial coproduct characteristic of the secret symmetry.

The presence of the hypercharge at level 0 would automatically extend the algebra, which transforms the scattering excitations, to $\alg{gl}(2|2)$, but this is not  straightforwardly compatible with the central extension. Nevertheless, recent progress \cite{Beisert:2014hya} based on the so-called $R\mathcal T\mathcal T$-formulation of the Yangian has revealed how the secret symmetry is non-trivially embedded in the algebra. However, it is still a challenge to obtain a translation to Drinfeld's picture of this discovery. Moreover, the issue of crossing symmetry is particularly delicate. 

It is also not clear how much of this secret symmetry is accidental to AdS${}_5$ and to the special features of the Lie superalgebra $\alg{psu}(2|2)$ \cite{Beisert:2006qh,Arutyunov:2009pw,Torrielli:2011zz}. The fact that a similar effect has been observed by now in a variety of ({\it a priori} unrelated) sectors of the AdS/CFT correspondence\footnote{This includes the boundary problem \cite{Regelskis:2011fa}, $n$-point amplitudes \cite{Beisert:2011pn}, pure-spinor formalism \cite{Berkovits:2011kn}, and quantum-affine deformations \cite{deLeeuw:2011fr}.} is an indication that this is not an isolated feature of the spectral problem. Nevertheless, all these other sectors still lie within AdS${}_5$ or close relatives. To see if this phenomenon is a truly universal feature of integrability within the AdS/CFT framework, it is crucial to extend the analysis beyond AdS${}_5$ to other dimensions.\footnote{The AdS${}_4$ background, for its nature of having a very similar S-matrix structure, does not seem to reveal much further on the secret symmetry issue.} At the same time, the deeper physical nature of the secret symmetry remains quite mysterious, and we hope that investigating it in other instances of the AdS/CFT correspondence might shed light on some of its most elusive properties. 

This is the direction we wish to pursue in this paper. Specifically, we find secret symmetries in the AdS${}_3$ spectral problem and analyse their features. Our present analysis also yields interesting information about the AdS${}_5$ case, when looking back in perspective, as we will explain next.

\paragraph{AdS${}_3$ integrable scattering.}
Recently, a new example of the AdS/CFT correspondence has become amenable to integrability methods. This involves type IIB backgrounds with an AdS${}_3$ factor in the metric, and the two most studied examples with 16 supersymmetries are $\AdS_3\times \Sphere^3\times \Torus^4$ and $\AdS_3\times \Sphere^3\times \Sphere^3\times \Sphere^1$. The latter background is characterised by a continuum parameter $\alpha$ related to the radii of the two 3-spheres, and which has a reflection in the appearance of the exceptional Lie superalgebra $\gen{D}(2,1;\alpha)\times \gen{D}(2,1;\alpha)$ as a superconformal algebra.\footnote{Recall from  \cite{Babichenko:2009dk} that the corresponding Metsaev--Tseytlin action \cite{Metsaev:1998it} is modelled on the supercoset space $[D(2,1;\alpha)\times D(2,1;\alpha)]/[SU(1,1)\times SU(2)\times SU(2)]$, where $D(2,1;\alpha)$ is the Lie supergroup corresponding to $\gen{D}(2,1;\alpha)$.}   Notice that in the contracting limit $\alpha \to 0$, the algebra reduces to $\alg{psu}(1,1|2)\times\alg{psu}(1,1|2)$, which, in turn, corresponds to the aforementioned background with the 4-torus factor. 

Such configurations provide instances of the $\AdS_3/\CFT_2$ correspondence. This was transferred into the framework of integrable systems by \cite{Babichenko:2009dk}, in which classical integrability was demonstrated (see also \cite{Sundin:2012gc}), a set of semiclassical finite gap equations for the spectrum were formulated, and a conjecture was put forward for an all-loop quantum Bethe ansatz. The latter was also fully elaborated into a spin chain picture \cite{OhlssonSax:2011ms}. The initial focus has been primarily on the massive BMN modes \cite{Berenstein:2002jq}, leaving aside the massless modes that now appear in contrast with the AdS${}_5$ case (see however \cite{Sax:2012jv,Lloyd:2013wza}). The problem of a fully consistent treatment of such massless modes has recently been investigated from a world-sheet perspective, paving the way to incorporate these excitations into the scaffolding of integrability \cite{Borsato:2014exa,Borsato:2014hja,Sfondrini:2014via}. Our motivation for pursuing an analysis of the quantum group symmetry algebra is dictated by the desire of eventually elucidating the role the massless modes in the full quantum group behind the scattering problem, building up on the Hopf algebra analysis of \cite{Borsato:2014hja}. One is interested in seeing if the algebra can help achieving the full description of the massless modes at the level of the universal $R$-matrix. We hope this paper's findings can provide a further step in that direction. 

To this end, we resort to the scattering theory developed in~\cite{Borsato:2012ud,Borsato:2012ss}. Specifically, the authors of \cite{Borsato:2012ud,Borsato:2012ss} derived an $S$-matrix and Bethe ansatz for the $\AdS_3\times \Sphere^3\times \Sphere^3\times \Sphere^1$ case from a  centrally-extended algebra of the Beisert type \cite{Beisert:2005tm,Arutyunov:2006ak}, adapted to a much smaller residual symmetry transforming the excitations. In a way, the smaller components of the algebra proliferate into several factors, only transferring the complication of a large single multiplet to many smaller ones instead. In particular, one has left and right components to worry about in the present case. We will describe this symmetry at length in the main text of the paper. 

In \cite{Borsato:2013qpa}, both the exact $S$-matrix and the Bethe ansatz for the $\AdS_3\times \Sphere^3\times \Torus^4$ case were constructed.\footnote{See also the earlier attempts \cite{David:2008yk,David:2010yg,Ahn:2012hw}.} This $S$-matrix received a series of confirmations from perturbative computations performed using the string sigma model \cite{Rughoonauth:2012qd,Sundin:2012gc,Abbott:2012dd,Beccaria:2012kb,Beccaria:2012pm,Sundin:2013ypa,Bianchi:2013nra,Engelund:2013fja,Bianchi:2014rfa}.  Suitable dressing phases to supplement the $S$-matrix and making it into a solution of the crossing equation were proposed in \cite{Borsato:2013hoa}. More studies have followed addressing various aspects of the spectral problem \cite{Abbott:2013ixa,Sundin:2013uca}. Furthermore, $\AdS_3\times \Sphere^3\times \Torus^4$ theories with mixed R-R and NS-NS flux have been analysed, and they provide in principle an interesting setup for the study of the conformal limit of a combined massive-massless integrable structure  \cite{Cagnazzo:2012se,Hoare:2013pma,Hoare:2013ida,Ahn:2014tua,David:2014qta,Babichenko:2014yaa}, cf.~\cite{Zamolodchikov:1992zr}. As a remark, we should say that it is quite crucial to test the dressing-phase solutions with new methods, and the algebraic one---via the universal $R$-matrix---could be a potential tool, as we will comment upon in the conclusions. The present treatment may therefore have a bearance on the problem of the dressing phases as well, as the inclusion of all symmetries of the system is essential to understand how to formulate the universal $R$-matrix for the complete centrally-extended algebra.\footnote{Let us remark that the universal $R$-matrix in \cite{Borsato:2013qpa} reproduces the matrix form of the individual $\alg{gl}(1|1)$ $S$-matrix blocks, and it is virtually insensitive to the central extension---were it not for the constraint on the representation parameters. It is still an open problem to find a global universal $R$-matrix  incorporating all the blocks and accounting for the central extension.}

Our algebraic treatment is general enough to encompass both backgrounds, hence we study the sphere case from the beginning. In fact, the only difference is that one is dealing with two copies of the fundamental $\alg{gl}(1|1)_l \times \alg{gl}(1|1)_r$ rather than only one copy, and having a parameter $s\in\mathbbm{R}$ ($s=\alpha$ for the modes we study in this paper) equal to $1$ in the torus case ($\alpha \to 1$). The two copies factorise anyway, and all our formul{\ae} just go through for arbitrary $s$ (we will only need to remember to keep $s$ non-zero, to remain inside the massive sector for the current concern of this paper). 

\subsection{Summary and outlook}

\paragraph{Summary.}
 In the first part of the paper, we find a complete realisation of the secret Yangian symmetries for $\AdS_3$ backgrounds, including their crossing symmetry condition. We will find two classes  of secret generators. One class is embedded in the $\alg{gl}(1|1)$ Yangian while the other one is not and at the same time more reminiscent of its higher-dimensional $\AdS_5$ analog. Due to this embedding (i.e.~the presence of a level-0 generator), we are able to relate these two classes by a quadratic map in the generators abstractly.\footnote{\label{fooqua} Because of their Yangian-like coproducts with the typical quadratic tail, a quadratic map is the most one would be allowed to have, to map not only the generators to each other but their coproducts as well.} For the {\it embedded} level-1 secret symmetry, an interpretation within the superconformal algebra of the theory (before the symmetry-breaking implemented by the choice of the spin-chain vacuum) is likely to occur in the $T^4$ case, where the level zero counterpart preserves the vacuum \cite{Borsato:2013qpa}. This is, however, not the case for the secret generator. By discussing the crossing symmetry relation of the latter, we gain a new perspective that might be useful in interpreting a recent observation of \cite{Beisert:2014hya}. In this paper, the secret symmetry was accommodated within crossing by allowing a shift in the multiplying parameter. Our results seem to suggest that even in the $\AdS_5$ case one might find it useful to think of the crossed secret generator as a {\it right} generator, to a {\it left} one being the secret symmetry originally found.

 In the second part of the paper, we find an incarnation of the $R\mathcal T\mathcal T$-construction of Beisert and de Leeuw's \cite{Beisert:2014hya} in the $\AdS_3$ case.  We see that we can reproduce the Yangian from the $R$-matrix via the $R\mathcal T\mathcal T$-construction, both for the left and the right scattering problem, extending the validity of their  framework to the lower-dimensional case at hand.

\paragraph{Outlook.}
Let us point out a few open problems which deserve further investigation, and serve as future direction of research.

Firstly, all our conclusions are extrapolated from specific representations. Therefore, one should study general representations to assess the universality of our results.

Secondly, the universal $R$-matrix for the complete centrally-extended algebra is still unknown. The same problem still plagues the $\AdS_5$ case, although the recent result of \cite{Beisert:2014hya} opened up the problem to a new promising approach. It has also been becoming clear that a significant re-interpretation of the Khoroshkin--Tolstoy formula has to occur for superalgebras with vanishing Killing form\footnote{We thank T. \L ukowski for discussions on this point.} \cite{Ferro:2012xw,Ale2}, and we hope that our analysis will provide further input to tackle the problem.

Finally, a world-sheet realisation of the secret symmetry in terms of non-local charges is still missing in any dimension, and we believe it is absolutely crucial to close this gap. This would be particularly important in view of the off-shell symmetry algebra approach of \cite{Arutyunov:2006ak,Borsato:2014exa}. We leave this fascinating problem to future investigations.

\section{Yangian of $\alg{gl}(1|1)_l\times \alg{gl}(1|1)_r$}\label{hopf-algebra}

In this section, we first review the features of Hopf superalgebras salient to the description of integrable scattering problems. For reviews, we refer the reader to \cite{Torrielli:2010kq,Torrielli:2011gg} and references therein. We then move on and summarise the relevant findings of \cite{Borsato:2013qpa} for the reader's convenience. Finally, we provide the details of the secret symmetries.

\subsection{Hopf superalgebra generalities}\label{hopf}

One first has to fix an algebra of symmetries of the system which one wishes to describe. In the superstring case, this turns out to be quite systematically a certain Lie superalgebra $\alg{g}$. To be able to deal with multiparticle states, one needs two additional maps which turn the algebra into a bi-superalgebra $A$. One is the {coproduct, $\Delta: A \rightarrow A \otimes A$, that encodes how the symmetry acts on two-particle states. The other map is the counit, $\epsilon: A \rightarrow \mathbbmss{C}$. A series of compatibility relations, most of which of immediate physical intuition, guarantee the consistency of the mathematical structure.

To go from a bi-superalgebra to a Hopf superalgebra, one equips the former with an antipode map, $\mathscr{S}: A \rightarrow A$, which is used to define the antiparticle (conjugated) representation to any given representation of the Lie superalgebra. The antipode is also subject to compatibility with the other maps, as we will have a chance to revisit in the main text. For its nature, the antipode is an anti-morphism, that is, $\mathscr{S} (ab) = (-)^{|a||b|} \mathscr{S}(b)\mathscr{S}(a)$ for all $a,b\in A$; here $|a|$ denotes the Gra{\ss}mann-parity of $a\in A$.  Moreover, one can prove that, if a bi-superalgebra has an antipode then the latter is unique.

To obtain the two-particle $S$-matrix, one first defines the action of symmetry on {\it in} states by means of the coproduct, as we stated above. On the other hand, the permuted map $P \circ \Delta =: \Delta^{\text{op}}$, with $P$ the graded permutation map, will be declared the action on {\it out} states. These two actions can differ, in general, extending standard textbook quantum mechanics where the Leibniz rule $\Delta (a) = a \otimes \mathbbmss{1} + \mathbbmss{1} \otimes a$ for all $a\in A$ guarantees the cocommutativity of the Hopf superalgebra, that is, $\Delta^{\text{op}} = \Delta$. Since, nevertheless, $\Delta$ and $\Delta^{\text{op}}$ generate tensor product representations with the same dimension, the two may be related by conjugation via an invertible element in the tensor product algebra, and this is the $S$-matrix or the $R$-matrix in mathematical literature:
\begin{equation}\label{eq:Rmatrixdef}
R\ \in\ A \otimes A\ewith \Delta^{\text{op}}(a)\,  R \ = \ R \, \Delta(a) \eforall a\ \in\ A~.
\end{equation} 
When this happens, the Hopf superalgebra is dubbed quasi-cocommutative, and, if the $R$-matrix satisfies an additional property which we will loosely relate to the physical bootstrap principle \cite{Zamolodchikov:1978xm,Dorey:1996gd}, it is also called quasi-triangular. The $S$-matrix needs to be compatible with the antipode, ensuring the physical crossing symmetry. A theorem of Drinfeld's shows that quasi-triangularity implies the Yang--Baxter equation and the crossing condition. 

Hopf superalgebras provide a particularly suitable framework to formulate integrable scattering problems. Moreover, they unify the treatment of arbitrary representations of the symmetry algebra (not only the fundamental particles but also the bound states transforming in higher irreducible representations) in one single language. The so-called universal $R$-matrix $\mathcal{R}$ (that is, the abstract solution to the quasi cocommutativity condition) has a special  importance and may sometimes be seen as an alternative to other approaches to the inverse scattering method.

\subsection{Lie algebra, representations, and $R$-matrices}
\label{Lie}

Let us begin by writing the action of the symmetry generators on the elementary scattering excitations, by focusing on one of the two copies of $\alg{gl}(1|1)_l\times \alg{gl}(1|1)_r$, say the left copy $\alg{gl}(1|1)_l$. The right copy can be studied in complete analogy \cite{Borsato:2013qpa}, and we will connect the two in section \ref{centro}. Each of these copies is further split into left and right representations that are related by crossing symmetry. The non-vanishing (anti-)commutation relations of the generators\footnote{For the sake of brevity, we shall omit the subscript `$l$' on the generators.} of $\alg{gl}(1|1)_l:=\langle\gen{B},\gen{H},\gen{Q},\gen{S}\rangle$ read as
\begin{equation}
[\gen{B}, \gen{Q}] \ = \ - 2 \gen{Q}~, \quad [\gen{B}, \gen{S}] \ = \ 2 \gen{S}~,\eand  \{\gen{Q}, \gen{S}\} \ = \ - \gen{H}~. 
\end{equation}
Here, $\gen{B}$ and $\gen{H}$ are Gra{\ss}mann-parity even (bosonic) and $\gen{Q}$ and $\gen{S}$ are Gra{\ss}mann-parity odd (fermionic), respectively. Moreover, $[\cdot,\cdot]$ denotes the commutator and  $\{\cdot,\cdot\}$ the anti-commutator. We will directly put ourselves into what was dubbed the {\it most symmetric frame} in \cite{Borsato:2013qpa}. The coproduct is obtained as (cf.~\cite{Gomez:2006va,Plefka:2006ze} for the AdS${}_5$ case)\footnote{We have included the unit element $\mathbbmss{1}$ in the description of the coproduct, since we will be dealing with the universal enveloping algebra (formed by arbitrary products and powers of the basic algebra generators) which has a unit. The element $\de^{\di p}$ is central.}
\begin{equation}
  \label{coprod}
  \begin{aligned}
    \Delta(\gen{B})\ :=\ \gen{B} \otimes \mathbbmss{1} + \mathbbmss{1} \otimes \gen{B}\eand&
    \Delta(\gen{H})\ :=\ \gen{H} \otimes \mathbbmss{1} + \mathbbmss{1} \otimes \gen{H}~, \\
     \Delta(\gen{Q})\ :=\ \gen{Q} \otimes \de^{-\di \frac{p}{4}} + \de^{\di \frac{p}{4}} \otimes \gen{Q}\eand&
     \Delta(\gen{S})\ :=\ \gen{S} \otimes \de^{\di \frac{p}{4}} + \de^{- \di \frac{p}{4}} \otimes \gen{S}~,\\
      \Delta(\de^{\di p})\ :=\ \de^{\di p} \otimes  \de^{\di p}\eand&
      \Delta(\mathbbmss{1})\ :=\ \mathbbmss{1} \otimes \mathbbmss{1}~, \qquad \qquad 
  \end{aligned}
\end{equation}
where $\di:=\sqrt{-1}$ and $p\in\mathbbm{R}$. Due to the centrality of $\de^{\di p}$, this coproduct is a Lie superalgebra homomorphism.\footnote{See also \cite{Hoare:2013pma,Hoare:2011fj}.} 

\paragraph{Left representation.}
The left representation of $\alg{gl}(1|1)_l$ is described by a left doublet $(\ket{\phi}, \ket{\psi})$ with symmetry action given by
\begin{subequations}
\begin{equation}\label{leftrep}
\begin{gathered}
\gen{B}_{\smallL} \ := \ \begin{pmatrix}1&0\\0&-1\end{pmatrix}~\qquad
\gen{H}_{\smallL} \ := \ {- \gamma^2\frac{h}{2}} \, \begin{pmatrix}1&0\\0&1\end{pmatrix},\\
\gen{Q}_{\smallL} \ := \ \gamma \sqrt{\frac{h}{2}}\begin{pmatrix}0&0\\1&0\end{pmatrix}~,\qquad
\gen{S}_{\smallL} \ := \ \gamma \sqrt{\frac{h}{2}}\begin{pmatrix}0&1\\0&0\end{pmatrix}, 
\end{gathered}
\end{equation}
where 
\begin{equation}\label{eq:defofex}
{\gamma\ :=\ \sqrt{\di (x^- - x^+)}}~, \quad  \frac{2 \di s}{h}\ =:\ x^+ + \frac{1}{x^+} -  x^- - \frac{1}{x^-} ~, \eand \de^{\di p} \ = \ \frac{x^+}{x^-}~,
\end{equation}
\end{subequations}
with $x^\pm\in\mathbbm{C}$ and $s,h\in\mathbbm{R}$.\footnote{The choice of a branch for the square root in $\gamma$ is not essential for the algebraic purposes of this paper, but it would matter for discussions on the analytic properties of the dressing phases.} Notice that $h$ is a function of the 't Hooft coupling. 
There is no need to specify the momentum generator $p$ as being left or right, as it will be common to the two representations. Letting $\Phi_{\smallL\smallL}\in\mathbbm{C}$ be an overall scalar factor (determined shortly), it can be checked that the left-left $R$-matrix denoted by  $R_{\smallL\smallL}$ and defined by
\begin{equation}\label{eq:RLL}
  \begin{aligned}
    R_{\smallL\smallL} (\ket{\phi} \otimes \ket{\phi})\ &:=\ {\Phi_{\smallL\smallL}} \, \frac{x_2^+ - x_1^-}{x_2^- - x_1^+} e^{i \frac{(p_1 - p_2)}{4}} \ket{\phi} \otimes \ket{\phi}, \\
    R_{\smallL\smallL} (\ket{\phi} \otimes \ket{\psi})\ &:=\ {\Phi_{\smallL\smallL}} \, \frac{x_2^+ - x_1^+}{x_2^- - x_1^+} e^{- i \frac{(p_1 + p_2)}{4}} \ket{\phi} \otimes \ket{\psi} + {\Phi_{\smallL\smallL}} \, \frac{x_2^+ - x_2^-}{x_2^- - x_1^+} \frac{\gamma_1}{\gamma_2} \ket{\psi} \otimes \ket{\phi}, \\
    R_{\smallL\smallL} (\ket{\psi} \otimes \ket{\phi})\ &:=\ {\Phi_{\smallL\smallL}} \, \frac{x_2^- - x_1^-}{x_2^- - x_1^+} e^{i \frac{(p_1 + p_2)}{4}} \ket{\psi} \otimes \ket{\phi} + {\Phi_{\smallL\smallL}} \, \frac{x_1^+ - x_1^-}{x_2^- - x_1^+} \frac{\gamma_2}{\gamma_1} \ket{\phi} \otimes \ket{\psi}, \\
    R_{\smallL\smallL} (\ket{\psi} \otimes \ket{\psi})\ &:=\ {\Phi_{\smallL\smallL}} \, \de^{i \frac{(p_2 - p_1)}{4}} \ket{\psi} \otimes \ket{\psi},
\end{aligned}
\end{equation}
where the indices $1$ and $2$ refer to the two scattering particles (first and second factor of the tensor product), indeed satisfies \eqref{eq:Rmatrixdef}, that is,
\begin{equation}\label{definiLL}
  \Delta^{\text{op}}_{\smallL\smallL} (\mathfrak{J})\,  R_{\smallL\smallL}\ =\ R_{\smallL\smallL}\, \Delta_{\smallL\smallL} (\mathfrak{J}) \eforall \mathfrak{J}\ \in\ \alg{gl}(1|1)_l~.
\end{equation}
The subscript in $\Delta^{\text{op}}_{\smallL\smallL}$ and $\Delta_{\smallL\smallL}$ means taking both factors of the coproduct (\ref{coprod}) in the left representation (\ref{leftrep}).

\paragraph{Right representation.}
Crossing symmetry relates left and right representations of $\alg{gl}(1|1)_l$. The action of the right representation on the right doublet of excitations $(\ket{\bar{\phi}},\ket{\bar{\psi}})$ is described by
\begin{equation}\label{rightrep}
\begin{gathered}
\gen{B}_{\smallR} \ := \ -\begin{pmatrix}1&0\\0&-1\end{pmatrix}~,\qquad
\gen{H}_{\smallR} \ := \ { -\frac{\gamma^2}{x^+x^-}\frac{h}{2} } \begin{pmatrix}1&0\\0&1\end{pmatrix},\\
\gen{Q}_{\smallR} \ := \ -\frac{\gamma}{\sqrt{x^+ \, x^-}} \sqrt{\frac{h}{2}}\begin{pmatrix}0&1\\0&0\end{pmatrix}~,\qquad
\gen{S}_{\smallR} \ := \ -\frac{\gamma}{\sqrt{x^+ \, x^-}} \sqrt{\frac{h}{2}}\begin{pmatrix}0&0\\1&0\end{pmatrix}
\end{gathered}
\end{equation}
with $\gamma$ and $h$ as in \eqref{eq:defofex}. The antipode $\mathscr{S}$ performing the connection is easily found applying the defining rule
\begin{equation}
  \label{rule}
  \mu \circ (\mathscr{S} \otimes \matId) \circ \Delta\ =\ \eta \circ \epsilon
\end{equation}
to the coproduct (\ref{coprod}); the map $\mu$ multiplies together two generators of the symmetry algebra, while $\epsilon$ is the counit and $\eta$ the unit map. Both $\eta$ and $\epsilon$ have to satisfy certain Hopf algebra consistency conditions with the multiplication and the coproduct map. In our case, these conditions amount to
\begin{equation}
\epsilon (\mathfrak{J})\ =\ 0\eforall\mathfrak{J}\ \in\ \alg{gl}(1|1)_l\eand\epsilon (\matId) \ =\ 1~,
\end{equation} 
and the antipode acts simply as
\begin{equation}
\mathscr{S} (\gen{J}) \ =\  - \gen{J}\eforall\mathfrak{J}\ \in\ \alg{gl}(1|1)_l~, \quad \mathscr{S} (\de^{\di p})\ =\ \de^{-\di p}~,\eand \mathscr{S} (\matId) \ =\  \matId~.
\end{equation}
Hence, the antipode map is idempotent and thus the same as its inverse. It was found in \cite{Borsato:2013qpa} that the left-right relation for any generator $\gen{J}$ can be written as
\begin{subequations}
\begin{equation} 
  \label{eq:underl}
  \mathscr{S} \big(\gen{J}_{\smallL}(x^\pm)\big) \ =\  \mathscr{C}^{-1} \bigg[\gen{J}_{\smallR}\bigg(\frac{1}{x^\pm}\bigg)\bigg]^{\str} \mathscr{C}~,
\end{equation}
where $\mathscr{C}$ is the matrix of charge conjugation
\begin{equation}
  \label{eq:C}
\mathscr{C}\ :=\ \begin{pmatrix}1&0\\0&\di\end{pmatrix},
\end{equation} 
\end{subequations}
and the apex ${}^{\str}$ denotes supertransposition. Charge conjugation allows to convert the left moving basis states into the right moving ones by means of $\ket{\phi}\mapsto\ket{\bar\phi}$ and $\ket{\psi}\mapsto\di\ket{\bar\psi}$.

Letting $\Phi_{\smallR\smallR}\in\mathbbm{C}$ be an overall scalar factor (determined shortly), it can be checked that the right-right $R$-matrix denoted by  $R_{\smallR\smallR}$ and defined by
\begin{equation}
\begin{aligned}
    R_{\smallR\smallR} (\ket{\bar{\phi}} \otimes \ket{\bar{\phi}})\ &:=\ {\Phi_{\smallR\smallR}} \frac{x_2^+ - x_1^-}{x_2^- - x_1^+} \de^{3 \di \frac{(p_1-p_2)}{4}} \ket{\bar{\phi}} \otimes \ket{\bar{\phi}}~,  \\
    R_{\smallR\smallR} (\ket{\bar{\phi}} \otimes \ket{\bar{\psi}})\ &:=\ {\Phi_{\smallR\smallR}} \frac{x_2^+ - x_1^+}{x_2^- - x_1^+}\de^{\di \frac{(p_1-3p_2)}{4}}\ket{\bar{\phi}} \otimes \ket{\bar{\psi}} + {\Phi_{\smallR\smallR}} \frac{\di  \gamma_1 \gamma_2 }{x_2^- - x_1^+}\de^{\di \frac{(p_1-p_2)}{2}} \ket{\bar{\psi}} \otimes \ket{\bar{\phi}}~, \\
    R_{\smallR\smallR} (\ket{\bar{\psi}} \otimes \ket{\bar{\phi}})\ &:=\ {\Phi_{\smallR\smallR}} \frac{x_2^- - x_1^-}{x_2^- - x_1^+} \de^{\di \frac{(3p_1-p_2)}{4}} \ket{\bar{\psi}} \otimes \ket{\bar{\phi}} + {\Phi_{\smallR\smallR}} \frac{\di \gamma_1 \gamma_2}{x_2^- - x_1^+}  \de^{\di \frac{(p_1-p_2)}{2}}\ket{\bar{\phi}} \otimes \ket{\bar{\psi}}~,  \\
    R_{\smallR\smallR} (\ket{\bar{\psi}} \otimes \ket{\bar{\psi}})\ &:=\ {\Phi_{\smallR\smallR}} \, \de^{\di \frac{(p_1-p_2)}{4}} \ket{\bar{\psi}} \otimes \ket{\bar{\psi}}
    \end{aligned}
    \end{equation}
    satisfies the right-version of   \eqref{eq:Rmatrixdef}.

\paragraph{Mixed representations.}
One can now project the coproduct (\ref{coprod}) into a mixed combination of right representation in the first factor and left representation in the second, namely $\Delta_{\smallR\smallL}$, and solve the equation for the scattering of a left mover with a right mover:
\begin{equation}
\label{definiRL}
  \Delta^{\text{op}}_{\smallR\smallL} (\mathfrak{J}) \,  R_{\smallR\smallL}\ =\ R_{\smallR\smallL}  \, \Delta_{\smallR\smallL} (\mathfrak{J})\eforall\mathfrak{J}\ \in\ \alg{gl}(1|1)_l~.
\end{equation}
Letting $\Phi_{\smallR\smallL}\in\mathbbm{C}$ be an overall scalar factor (determined shortly), the corresponding $R$-matrix reads
\begin{equation}\label{eq:RRL}
\begin{aligned}
    R_{\smallR\smallL} (\ket{\bar{\phi}} \otimes \ket{\phi})\ &:=\ {\Phi_{\smallR\smallL}} \, \frac{x_2^- x_1^+ - 1}{x_2^+ x_1^+ - 1} \de^{\di \frac{(p_1+p_2)}{4}} \ket{\bar{\phi}} \otimes \ket{\phi} + {\Phi_{\smallR\smallL}} \, \frac{\di \gamma_1 \gamma_2}{(x_2^+ x_1^+ - 1)} \de^{\di\frac{p_1}{2}} \ket{\bar{\psi}} \otimes \ket{\psi}~,  \\
    R_{\smallR\smallL} (\ket{\bar{\phi}} \otimes \ket{\psi})\ &:=\ {\Phi_{\smallR\smallL}} \, \frac{x_2^- x_1^- - 1}{x_2^+ x_1^+ - 1} \de^{\di \frac{(3 p_1 + p_2)}{4}} \ket{\bar{\phi}} \otimes \ket{\psi}~, \\
    R_{\smallR\smallL} (\ket{\bar{\psi}} \otimes \ket{\phi})\ &:=\ {\Phi_{\smallR\smallL}} \de^{\di\frac{(p_1-p_2)}{4}} \ket{\bar{\psi}} \otimes \ket{\phi}~, \\
    R_{\smallR\smallL} (\ket{\bar{\psi}} \otimes \ket{\psi})\ &:=\ {\Phi_{\smallR\smallL}} \, \frac{x_2^+ x_1^- - 1}{x_2^+ x_1^+ - 1} \de^{\di \frac{(3 p_1- p_2)}{4}} \ket{\bar{\psi}} \otimes \ket{\psi} + {\Phi_{\smallR\smallL}} \, \frac{\di  \gamma_1 \gamma_2}{x_2^+ x_1^+ - 1} \de^{i\frac{p_1}{2}} \ket{\bar{\phi}} \otimes \ket{\phi}~,
    \end{aligned}
\end{equation}
and satisfies the crossing equation\footnote{This equation is derived from the Hopf superalgebra universal relation $(\mathscr{S} \otimes \matId) \mathcal R = \mathcal R^{-1}= (\matId \otimes \mathscr{S}^{-1}) \mathcal R$.} (cf.~\cite{Janik:2006dc})
\begin{equation}
  \label{matricial}
  (\mathscr{C}^{-1} \otimes \matId) R_{\smallR\smallL}^{\, \, \str_1}\Big(\frac{1}{x_1^\pm}, x_2^\pm\Big) (\mathscr{C} \otimes \matId) R_{\smallL\smallL} (x_1^\pm, x_2^\pm) 
  \ =\ \matId \otimes \matId
\end{equation}
($\str_i$ meaning supertransposition in the factor $i$), provided the scalar factors are related by a specific condition. Such a condition was the object of the analysis in \cite{Borsato:2013hoa}.

The other possible combination of mixed scattering is obtained in \cite{Borsato:2013qpa} by solving the analog of the coproduct relations (\ref{definiLL}) and (\ref{definiRL}) projected into the appropriate representations, the result being
\begin{equation}\label{eq:RLR}
\begin{aligned}
    R_{\smallL\smallR} (\ket{\phi} \otimes \ket{\bar{\phi}})\ &:=\ {\Phi_{\smallL\smallR}} \frac{x_2^- x_1^+ - 1}{x_2^- x_1^- - 1} \de^{-\di \frac{(p_1+p_2)}{4}} \ket{\phi} \otimes \ket{\bar{\phi}} +  {\Phi_{\smallL\smallR}} \frac{\di \gamma_1 \gamma_2}{x_2^- x_1^- - 1} \de^{-\di \frac{p_2}{2}} \ket{\psi} \otimes \ket{\bar{\psi}}~,\\
    R_{\smallL\smallR} (\ket{\phi} \otimes \ket{\bar{\psi}})\ &:=\ {\Phi_{\smallL\smallR}} \, \de^{\di \frac{(p_1-p_2)}{4}} \ket{\phi} \otimes \ket{\bar{\psi}}~, \\
    R_{\smallL\smallR} (\ket{\psi} \otimes \ket{\bar{\phi}})\ &:=\ {\Phi_{\smallL\smallR}} \frac{x_1^+ x_2^+ - 1}{x_1^- x_2^- - 1} \de^{- \di \frac{(p_1+3p_2)}{4}} \ket{\psi} \otimes \ket{\bar{\phi}},\\
    R_{\smallL\smallR} (\ket{\psi} \otimes \ket{\bar{\psi}})\ &:=\ {\Phi_{\smallL\smallR}} \frac{x_2^+ x_1^- - 1}{x_2^- x_1^- - 1} \de^{\di \frac{(p_1-3p_2)}{4}} \ket{\psi} \otimes \ket{\bar{\psi}} + {\Phi_{\smallL\smallR}} \frac{\di  \gamma_1 \gamma_2}{x_2^- x_1^- - 1}  \de^{- \di \frac{p_2}{2}} \, \ket{\phi} \otimes \ket{\bar{\phi}}~,
    \end{aligned}
\end{equation}
where $\Phi_{\smallL\smallR}\in\mathbbm{C}$ is an overall scalar factor (determined shortly).

The remaining crossing equations analogous to \eqref{matricial}, and similarly dealt with in \cite{Borsato:2013hoa}, read 
\begin{equation}
\begin{aligned}
  (\mathscr{C}^{-1} \otimes \matId) \, R_{\smallR\smallR}^{\, \, \str_1}\Big(\frac{1}{x_1^\pm}, x_2^\pm\Big) \, (\mathscr{C} \otimes \matId) \, R_{\smallL\smallR} (x_1^\pm, x_2^\pm) \ &=\ \matId \otimes \matId~,\\
  (\matId \otimes \mathscr{C}^{-1}) \, R_{\smallL\smallR}^{\, \, \str_2}\Big(x_1^\pm, \frac{1}{x_2^\pm}\Big) \, (\matId \otimes \mathscr{C} ) \, R_{\smallL\smallL} (x_1^\pm, x_2^\pm)\ &=\ \matId \otimes \matId~,\\
(\matId \otimes \mathscr{C}^{-1}) \, R_{\smallR\smallR}^{\, \, \str_2}\Big(x_1^\pm, \frac{1}{x_2^\pm}\Big) \, (\matId \otimes \mathscr{C} ) \, R_{\smallR\smallL} (x_1^\pm, x_2^\pm)\ &=\  \matId \otimes \matId~.
\end{aligned}
\end{equation}

Let us notice the the above $R$-matrices have the following property:
\begin{eqnarray}
R_{\smallR\smallR}\ =\ \frac{\Phi_{\smallR\smallR}}{\Phi_{\smallL\smallL}} \, \de^{\di \frac{(p_1 - p_2)}{2}} R_{\smallL\smallL}\eand
R_{\smallL\smallR} \ =\  \frac{\Phi_{\smallL\smallR}}{\Phi_{\smallR\smallL}} \, \de^{-\di \frac{(p_1 + p_2)}{2}} \, \frac{x_2^+ x_1^+ - 1}{x_2^- x_1^- - 1} \, R_{\smallR\smallL}~.
\end{eqnarray}

\paragraph{Scalar factors.}
We report for completeness the equations imposed on the scalar factors by the crossing equations, together with the requirements of unitarity\footnote{In what follows, $R_{21} = \big[P\circ R\big](x_2^\pm, x_1^\pm) $ when we deal with $R$-matrices, while $\Phi_{21} = \Phi(x_2^\pm, x_1^\pm)$ when we deal with scalar factors.}
\begin{equation}
\begin{gathered}
[R_{\smallL\smallL}]_{12} \, [R_{\smallL\smallL}]_{21}\ =\ \mathbbmss{1} \otimes \mathbbmss{1}~, \quad [R_{\smallR\smallL}]_{12} \, [R_{\smallL\smallR}]_{21}\ =\ \mathbbmss{1} \otimes \mathbbmss{1}~,\\
[R_{\smallL\smallR}]_{12} \, [R_{\smallR\smallL}]_{21}\ =\ \mathbbmss{1} \otimes \mathbbmss{1}~, \quad [R_{\smallR\smallR}]_{12} \, [R_{\smallR\smallR}]_{21}\ =\ \mathbbmss{1} \otimes \mathbbmss{1}~.
\end{gathered}
\end{equation}
We will not describe the solution proposed in \cite{Borsato:2013hoa}, since we will not need such a solution for the present algebraic purposes. The conditions are 
\begin{equation}
\begin{gathered}
  {\Phi_{\smallL\smallL}} \, [\Phi_{\smallR\smallL}]_{\bar{1}}\ =\ \frac{x_2^+ - x_1^+}{x_2^+ - x_1^-}~, \qquad {\Phi_{\smallL\smallR}} \, [\Phi_{\small\smallRR}]_{\bar{1}}\ =\ \frac{x_1^+ - \frac{1}{x_2^-}}{x_1^+ - \frac{1}{x_2^+}}~,\\
  {\Phi_{\smallL\smallL}} \, [\Phi_{\smallL\smallR}]_{\bar{2}}\ =\ \frac{x_2^- - x_1^-}{x_2^+ - x_1^-}~, \qquad 
  {\Phi_{\smallR\smallL}}\,  [\Phi_{\smallR\smallR}]_{\bar{2}}\ =\ \frac{x_2^- - \frac{1}{x_1^+}}{x_2^- - \frac{1}{x_1^-}}~,\\
  {\Phi_{\smallL\smallL}} \, [{\Phi_{\smallL\smallL}}]_{21}\ =\ 1~, \qquad 
  {\Phi_{\smallL\smallR}} \, [{\Phi_{\smallR\smallL}}]_{21}\ =\ 1~, \qquad 
  {\Phi_{\smallR\smallR}} \, [{\Phi_{\smallR\smallR}}]_{21}\ =\ 1~.
  \end{gathered}
\end{equation}
In the above, $[\Phi]_{\, \bar{i}}$ denotes the antiparticle map in the variable $i$, namely $x_i^\pm \mapsto 1/x_i^\pm$. 

\subsection{Yangian secret symmetries}\label{sec:YSS}

In this section, we display the $R$-matrix Yangian symmetry as found in \cite{Borsato:2013qpa}, completing the information that was missing there regarding the type $\gen{B}$ hypercharge generator at Yangian level 1. We shall focus on the Yangian $\mathcal Y(\mathfrak{gl}(1|1)_l)$ of $\mathfrak{gl}(1|1)_l$.

\paragraph{Secret symmetry I.}
It can be shown that the left representation of the level-1 Yangian generators of $\mathcal Y(\mathfrak{gl}(1|1)_l)$
\begin{subequations}
\begin{equation}\label{eq:LL1-YG}
{\gen{e}_{1}}_{\smallL} \ :=\ u_{\smallL} \, \gen{Q}_{\smallL}~, \quad 
{\gen{f}_{1}}_{\smallL} \ :=\ u_{\smallL} \, \gen{S}_{\smallL}~, \quad
{\gen{h}_{1}}_{\smallL} \ :=\ u_{\smallL} \, \gen{H}_{\smallL}~, \eand 
{\gen{b}_{1}}_{\smallL} \ :=\ u_{\smallL} \, \gen{B}_{\smallL}~,
\end{equation}
with the left spectral parameter
\begin{equation}\label{eq:uLspecpar}
u_{\smallL} \ := \ \di\tfrac{h}{2} \, x^+
\end{equation}
\end{subequations}
and the corresponding right representation of the level-1 Yangian generators of  $\mathcal Y(\mathfrak{gl}(1|1)_l)$
\begin{subequations}
\begin{equation}\label{eq:RL1-YG}
{\gen{e}_{1}}_{\smallR} \ :=\ u_{\smallR} \, \gen{Q}_{\smallR}~, 
\quad {\gen{f}_{1}}_{\smallR} \ :=\ u_{\smallR} \, \gen{S}_{\smallR}~,
\quad {\gen{h}_{1}}_{\smallR} \ :=\ u_{\smallR} \, \gen{H}_{\smallR}~, \eand 
{\gen{b}_{1}}_{\smallR} \ :=\ \begin{pmatrix}\omega_{11}&0\\0&\omega_{22}\end{pmatrix}
\end{equation}
with
\begin{equation}
\omega_{11}\ := \ 1 + \di\frac{h}{2 x^+} - \di\frac{h}{x^-}~, \quad \omega_{22}\ :=\ 1 + \di\frac{h}{2 x^+}~, \eand u_{\smallR} \ := \ \di\frac{h}{2 x^-}~,
\end{equation}
\end{subequations}
are symmetries of the $R$-matrices which we displayed in the previous sections; here $x^\pm$ and $h$ are as in \eqref{eq:defofex}. The novelty with respect to \cite{Borsato:2013qpa} is the expression for the right level-1 hypercharge.\footnote{Notice that the generators of type $\gen{Q}$, $\gen{S}$, and $\gen{H}$ form an ideal inside the Yangian, since type $\gen{B}$ is never produced by commuting any of the elements of the ideal. In view of this fact, in principle, one might like to disregard a multiple of the identity added to $\gen{b}_1$. However, this would not agree with the crossing symmetry to be described shortly.} These generators being symmetries amounts to saying that by projecting one and the same universal expression for each level-1 Yangian coproduct $\Delta(\gen{j}_1)$ each time in the appropriate combination of representations, one satisfies all the relations 
\begin{equation}
\Delta^{\text{op}}_{kl} ({\gen{j}_1}) \, R_{kl} \, = \, R_{kl} \, \Delta_{kl} ({\gen{j}_1}),
\end{equation}
with $k,l \in \{{\smallL},{\smallR}\}$ and $\gen{j}_1\in\{ \gen{e}_1, \gen{f}_1, \gen{h}_1, \gen{b}_1\}$.  After defining the level-0 generators simply as
\begin{equation}
{\gen{e}_{0}}_{\smallR} \ := \ \gen{Q}_{\smallR}~, \quad {\gen{f}_{0}}_{\smallR} \ := \ \gen{S}_{\smallR}~,\quad 
{\gen{h}_{0}}_{\smallR} \ := \ \gen{H}_{\smallR}~,\eand {\gen{b}_{0}}_{\smallR} \ := \ \gen{B}_{\smallR}~,
\end{equation}
the coproducts for the generators of the $\alg{sl}(1|1)$ ideal, already given in \cite{Borsato:2013qpa}, look like
\begin{subequations}
\begin{equation}\label{eq:random1}
  \begin{aligned}
    \Delta({\gen{e}}_1)\ &:=\ \gen{e}_1 \otimes \de^{-\di \frac{p}{4}} + \de^{\di \frac{p}{4}} \otimes \gen{e}_1 + \de^{\di \frac{p}{4}}\,\gen{h}_0  \otimes \gen{e}_0~, \\
    \Delta(\gen{f}_1)\ &:=\ \gen{f}_1 \otimes \de^{\di \frac{p}{4}} + \de^{-\di \frac{p}{4}} \otimes \gen{f}_1 + \gen{f}_0 \otimes e^{i \frac{p}{4}} \, \gen{h}_0~, \\
    \Delta(\gen{h}_1)\ &:=\ \gen{h}_1 \otimes \matId + \matId \otimes \gen{h}_1 + \gen{h}_0 \otimes \gen{h}_0~,
  \end{aligned}
\end{equation}
where we have here made explicit the braiding by the $\de^{\di p}$-type central element in the frame we are using. The novelty is that we now have a level-1 canonical coproduct for the hypercharge given by
\begin{equation}\label{eq:random2}
\Delta({\gen{b}}_1)\ :=\ \gen{b}_1 \otimes \mathbbmss{1} + \mathbbmss{1} \otimes \gen{b}_1 \, - 2 \, \gen{f}_0 \, \de^{\di \frac{p}{4}} \otimes \gen{e}_0 \, \de^{\di \frac{p}{4}} \, + \, \gen{b}_0 \otimes \gen{b}_0~.
\end{equation} 
\end{subequations}
With these formul{\ae}, one can prove that the level-0 and level-1 generators (both in the left and the right representation), as well their coproducts (in all possible combinations of left and right choices), are compatible with the so-called Drinfeld's second realisation \cite{Drinfeld:1987sy} of the $\alg{gl}(1|1)$ Yangian:  
\begin{equation}
  \label{eq:Lie}
  \begin{gathered}
    \comm{\gen{b}_0}{\gen{e}_n}\ =\ -2 \, \gen{e}_n~, \qquad
    \comm{\gen{b}_0}{\gen{f}_n}\ =\ 2 \, \gen{f}_n~, \qquad
    \acomm{\gen{e}_m}{\gen{f}_n}\ =\ -\gen{h}_{m+n}~, \\
    \comm{\gen{b}_m}{\gen{b}_n}\ =\ 0~, \qquad \comm{\gen{h}_m}{\cdot}\ =\ 0~, \qquad 
    \acomm{\gen{e}_m}{\gen{e}_n}\ =\ 
    \acomm{\gen{f}_m}{\gen{f}_n}\ =\ 0~, \\
    \comm{\gen{b}_{m+1}}{\gen{e}_n} - \comm{\gen{b}_m}{\gen{e}_{n+1}} + \acomm{\gen{b}_m}{\Ygen{e}_n}\ =\ 0~, \qquad
    \comm{\gen{b}_{m+1}}{\gen{f}_n} - \comm{\gen{b}_m}{\gen{f}_{n+1}} - \acomm{\gen{b}_m}{\gen{f}_n}\ =\ 0~
  \end{gathered}
\end{equation}
for $m,n\in\mathbbm{N}_0$.

Furthermore, all these symmetries, including the hypercharge generator, satisfy the
crossing symmetry. This proceeds as already described in \cite{Borsato:2013qpa}. Firstly, one derives from the coproduct the expression for the antipode utilising \eqref{rule} (with $\epsilon$ annihilating all level-1 Yangian generators), obtaining
\begin{equation}
\begin{gathered}
\mathscr{S} (\gen{e}_1)\ =\ - \gen{e}_1 + \gen{e}_0 \, \gen{h}_0~, \qquad 
  \mathscr{S} (\gen{f}_1)\ =\ - \gen{f}_1 + \gen{f}_0 \, \gen{h}_0~, \qquad 
  \mathscr{S} (\gen{h}_1)\ =\ - \gen{h}_1 + \gen{h}_0^2~,\\
\mathscr{S} (\gen{b}_1)\ =\ - \gen{b}_1 - 2 \, \gen{f}_0 \, \gen{e}_0 + \gen{b}_0^2~.
\end{gathered}
\end{equation} 
At this point, one can verify that the equation 
\begin{equation} 
  \mathscr{S} \big({\gen{j}_1}_{\smallL}(x^\pm)\big)\ =\ \mathscr{C}^{-1} \bigg[{\gen{j}_1}_{\smallR}\bigg(\frac{1}{x^\pm}\bigg)\bigg]^{\str} \mathscr{C}~,
\end{equation}
indeed holds with the same charge conjugation matrix \eqref{eq:C}, and for all the generators including the hypercharge, that is, for ${\gen{e}_1}$, ${\gen{f}_1}$, ${\gen{h}_1}$, and $ {\gen{b}_1}$.

\paragraph{Secret symmetry II.}
In the previous paragraph, we have found a secret symmetry (hypercharge generator) which, in constract with the $\AdS_5$ case \cite{deLeeuw:2012jf,Spill:2008tp}, is embedded in Drinfeld's second realisation of the relevant $S$-matrix Yangian. In fact, there is a whole class $[\gen{b}_1]:= \gen{b}_1+s\mathbbm{1}$ for $s\in\mathbbm{R}$ of secret generators preserving all desired relations.\footnote{The coproduct on equivalence classes $[\gen{j}]:=\gen{j}+s\mathbbm{1}$ is consistently defined up to multiples of $\mathbbm{1}\otimes\mathbbm{1}$.} Next, we would like to show that, in the case of $\AdS_3$, there exists another class of secret symmetry generators, not embedded in Drinfeld's second realisation of the Yangian, and related to the class $[\gen{b}_1]$ by performing a certain quadratic\footnote{As anticipated in footnote \ref{fooqua}, a quadratic transformation like \eqref{mappa} is the only one that has a chance of mapping not only the generators, but also their coproducts.}  Drinfeld map of the form
\begin{eqnarray}
\label{mappa}
[\widehat{\gen{j}}]\ :=\ \widehat{\gen{j}}+t\mathbbm{1}\efor t\ \in\ \mathbbm{R}\ewith [\widehat{\gen{j}}]\  =\ c\,  [\gen{j}_1]  + c_{AB} \, \gen{j}_0^A \, \gen{j}_0^B~.
\end{eqnarray}
Here, $\widehat{\gen{j}}$ is the level-1 Yangian generator in Drinfeld's first realisation that is associated with $\gen{j}_1$, $(c, c_{AB})$ are some constant coefficients, and $\gen{j}_0^A$ are the level-0 generators including $\gen{b}_0$. Such map is used to switch between Drinfeld's first and second realisations.

The generator we will now present is a symmetry of the $R$-matrix in all possible combinations of left and right representations, and, similarly to the hypercharge in the previous section, satisfies crossing symmetry. The expression for this alternative secret symmetry (which we call $\widehat{\gen{b}}$) is much closer to its AdS${}_5$ analog, and it may be the true AdS${}_3$ correspondent of that general phenomenon.

The symmetry is given by
\begin{subequations}
\begin{equation}\label{eq:2ndSecSym}
\widehat{\gen{b}}_{\smallL} \ :=\ \delta \, \gen{B}_{\smallL} \eand \widehat{\gen{b}}_{\smallR} \ :=\ \begin{pmatrix}\tau_{11}&0\\0&\tau_{22}\end{pmatrix},
\end{equation}
where
\begin{equation}
\quad \delta\ :=\ - \di\frac{h}{4} \, (x^+ + x^-)~,\quad
\tau_{11}\ :=\ \di\frac{h}{4}\bigg(\frac{3}{x^-} - \frac{1}{x^+} \bigg)\,,\eand 
\tau_{22}\ :=\ - \di\frac{h}{4}\bigg(\frac{3}{x^+} - \frac{1}{x^-} \bigg)
\end{equation}
\end{subequations}
and $x^\pm$ and $h$ as in \eqref{eq:defofex}. Its universal coproduct reads
\begin{equation}\label{eq:2ndSecSymCoProd}
\Delta(\widehat{\gen{b}})\ :=\  \widehat{\gen{b}} \otimes \mathbbmss{1} + \mathbbmss{1} \otimes \widehat{\gen{b}} + \, \gen{e}_0 \, \de^{-\di \frac{p}{4}} \otimes \gen{f}_0 \, \de^{-\di \frac{p}{4}}\, + \, \gen{f}_0 \, \de^{\di \frac{p}{4}} \otimes \gen{e}_0 \, \de^{\di \frac{p}{4}}~,
\end{equation} 
implying an antipode
\begin{equation}
\label{analogo}
\mathscr{S} (\widehat{\gen{b}})\ =\ - \widehat{\gen{b}} - \gen{h}_0~.
\end{equation}
Crossing reads the same way as for all the other generators we have in this paper, that is,
\begin{equation} 
  \label{crossBhat}
  \mathscr{S} \big(\widehat{\gen{b}}_{\smallL}(x^\pm)\big)\ =\ \mathscr{C}^{-1} \bigg[\widehat{\gen{b}}_{\smallR}\bigg(\frac{1}{x^\pm}\bigg)\bigg]^{\str} \mathscr{C}~,
\end{equation}
where $\mathscr{C}$ was given in \eqref{eq:C}. Finally, the quadratic map \eqref{mappa} is given by
\begin{equation}
 [\widehat{\gen{b}}]\ =\ -[\gen{b}_1]+\tfrac12(\gen{e}_0\gen{f}_0-\gen{f}_0\gen{e}_0+\gen{b}_0\gen{b}_0)~.
\end{equation}
The relation between the parameters $s$ and $t$ in $[\gen{b}_1]=\gen{b}_1+s\mathbbm{1}$ and $[\widehat{\gen{b}}]=\widehat{\gen{b}}+t\mathbbm{1}$ is fixed by the choice of matrix representation of the generators. Note that we are making explicit use of the level-0 generator $\gen{b}_0$ to establish the relation between the two classes of secret generators.  Because of this, it remains to be seen whether  the two secret  symmetries can, in fact, be identified. We emphasise that this is in contrast with the AdS$_5$ case where no such embedding exists.

\paragraph{$\AdS_5$ versus $\AdS_3$ secret symmetries.}
Our above analysis in the $\AdS_3$ case yields a new perspective on the $\AdS_5$ problem. In particular,
our analysis provides a re-interpretation of the observation made in \cite{Beisert:2014hya} concerning how to incorporate the secret symmetry within crossing (a problem that was observed in \cite{Matsumoto:2007rh}). In fact, even in the AdS${}_5$ case we may accept that the crossed generator is some sort of {\it right} generator, to a {\it left} one being the secret symmetry originally found. Most of the AdS${}_5$ magnon multiplet is crossing self-dual, apart from the secret generator, which indeed satifies a formula perfectly analogous to (\ref{analogo}). Therefore, as noticed in \cite{Beisert:2014hya}, the secret symmetry respects (\ref{crossBhat}) by defining
(in the notation of \cite{Matsumoto:2007rh})
\begin{equation}
\gen{J}_R (\mbox{AdS}_5) \ = \ \bigg[- \gen{J} (\mbox{AdS}_5) + \frac{2\di}{g} \gen{C}\bigg]_{x^\pm\mapsto 1/x^\pm}~,
\end{equation}
where $\gen{C}$ is the central charge associated to the magnon energy, and $g$ is the square root of the AdS${}_5$ 't Hooft coupling divided by $4 \pi$.

\section{$R\mathcal T\mathcal T$-realisation for the deformed $\mathfrak{gl}(1|1)_l\times\mathfrak{gl}(1|1)_r$ Yangian}

In this section, we construct the $R\mathcal T\mathcal T$-realisation for the Yangian $\mathcal Y(\mathfrak{gl}(1|1)_l\times\mathfrak{gl}(1|1)_r)$ originating from the full  $R$-matrix for $\AdS_3\times S^3\times M^4$.  After a brief review in which we shall explain how the generators arise directly from the $R$-matrix, we spell out some of the generators explicitly. Eventually, we discuss central extensions. Our discussion follows the ideas of \cite{Beisert:2014hya} to which we refer for further details of the construction.

\subsection{$R\mathcal T\mathcal T$-generalities}

The following discussion is tailored to the Lie superalgebra $\alg{gl}(p|q)$. In particular, let $V$ be a Hilbert superspace with  $\text{dim}_{\mathbbm{C}}(V)=p|q$. Furthermore, let 
\begin{equation}
R\,:\,\mathbbm{C}^2\ \to\ \text{End}(V\otimes V)
\end{equation}
be the $R$-matrix which is  invariant under the action of the Yangian $\mathcal Y(\alg{gl}(p|q)):=\langle \gen{J}_m{}_\bB{}^\bA\rangle$, where $\bA=(A,I)$, $\bB=(B,J)$, $\ldots$ with $A,B,\ldots=1,\ldots,p$ and $I,J,\ldots=p+1,\ldots,p+q$, for a specific choice of representations. This $R$-matrix is unique up to scalar factors, and it is typically a rational function of two spectral parameters $(u, v)\in\mathbbm{C}^2$. Here, $m\in\mathbbm{N}_0$ denotes the Yangian level in a suitable basis. Moreover, we let $|\bA|$ denote the Gra{\ss}mann-parity of $\bA=(A,I)$, that is, $|A|=0$ and $|I|=1$. For instance, the defining (anti-)commutation relations of $\mathcal Y (\alg{gl}(p|q))$ are{
\begin{equation}
\begin{aligned}
\gcomm{\gen{J}_0{}_\bB{}^\bA,\gen{J}_0{}_\bD{}^\bC}\ &=\ (-)^{|\bB|}\delta_\bB{}^\bC\gen{J}_0{}_\bD{}^\bA-(-)^{|\bB||\bC|+|\bB||\bD|+|\bC||\bD|}\delta_\bD{}^{\bA}\gen{J}_0{}_\bB{}^\bC~,\\
\gcomm{\gen{J}_0{}_\bB{}^\bA,\gen{J}_1{}_\bD{}^\bC}\ &=\ (-)^{|\bB|}\delta_\bB{}^\bC\gen{J}_1{}_\bD{}^\bA-(-)^{|\bB||\bC|+|\bB||\bD|+|\bC||\bD|}\delta_\bD{}^{\bA}\gen{J}_1{}_\bB{}^\bC~.
\end{aligned}
\end{equation}}

Next, letting 
\begin{equation}
\bb T_\bB{}^\bA\,:\,\mathbbm{C}\ \to\ \mathcal Y (\alg{gl}(p|q))
\end{equation}
be functions of a spectral parameter, say $u\in\mathbbm{C}$, holomorphic in a vicinity of $u=\infty$, and $\{E_\bB{}^\bA\}$ be the standard basis\footnote{That is, the only non-zero entry of $E_\bB{}^\bA$ is $(-)^{|\bB|}$ in row $\bA$ and column $\bB$.} for $\text{End}(V)$, we may define a map
\begin{equation}
\mathcal T\ :\ \mathbbm{C}\ \to\ \text{End}(V)\otimes \mathcal Y (\alg{gl}(p|q))\ewith
\mathcal T\ :=\ E_\bA{}^\bB\otimes \bb T_\bB{}^\bA~.
\end{equation}
Given this data, the Yangian $\mathcal Y(\alg{gl}(p|q))$ follows from the tensor algebra $\mathscr T(\braket{\bb T_\bB{}^\bA(u)})$ of the algebra $\braket{\bb T_\bB{}^\bA(u)}$ modulo the so-called $R\mathcal T\mathcal T$-relations
\begin{equation}\label{eq:RTTRel}
R_{12}(u,v)\mathcal T_{13}(u)\mathcal T_{23}(v)\ =\ \mathcal T_{23}(v)\mathcal T_{13}(u)R_{12}(u,v)~,
\end{equation}}
that is,
\begin{equation}
\mathcal Y(\alg{gl}(p|q))\ =\ \frac{\mathscr T({\braket{\bb T}})}{\braket{R_{12}\mathcal T_{13}\mathcal T_{23}-\mathcal T_{23}\mathcal T_{13}R_{12}}}~,
\end{equation}

Indeed, by expanding the $R\mathcal T\mathcal T$-relations \eqref{eq:RTTRel} around { $(u,v)=(\infty,\infty)$}, one can see that the Laurent coefficients $\bb T_m{}_\bB{}^\bA$ of  $\bb T_\bB{}^\bA$,
\begin{equation}
\bb T_\bB{}^\bA(u)\ =\ \sum_{m\geq 0} u^{-m}\,\bb T_{m-1}{}_\bB{}^\bA~,
\end{equation}
can be combined to form a new set of generators $\bb J_m{}_\bA{}^\bB$ which are in one-to-one correspondence with the Yangian generators $\gen{J}_m{}_\bA{}^\bB$. For instance, the $\bb J_m{}_\bB{}^\bA$s arising in this framework from the $RTT$-relations with the standard rational Yang-type $R$-matrix $R(u,v)=\mathbbmss{1} + \frac{P}{u-v}$ (with $P$ the graded permutation operator) are 
\begin{equation}\label{eq: frombbTtobbJ}
\begin{gathered}
\bb J_{-1}{}_\bB{}^\bA\ =\ \bb T_{-1}{}_\bB{}^\bA~,\qquad
\bb J_{0}{}_\bB{}^\bA\ =\ \bb T_{0}{}_\bB{}^\bA~,\\
\bb J_{1}{}_\bB{}^\bA\ =\ \bb T_{1}{}_\bB{}^\bA-\tfrac{1}{2}{(-)^{(|\bA|+|\bC|)(|\bB|+|\bC|)}}\bb T_{0}{}_\bB{}^\bC\bb T_{0}{}_\bC{}^\bA~,
\end{gathered}
\end{equation}
and progressively more complicated for $\bb J_{m}{}_\bB{}^\bA$ with $m>1$. One notices that whilst $\bb J_{-1}{}_\bB{}^\bA$ is central and hence must be proportional to the identity, the generators $\bb J_{0}{}_\bB{}^\bA$ and $\bb J_{1}{}_\bB{}^\bA$ obey \eqref{eq: frombbTtobbJ} provided $\gen{J}_{0}{}_\bB{}^\bA\leftrightarrow\bb J_{0}{}_\bB{}^\bA$ and $\gen{J}_{1}{}_\bB{}^\bA\leftrightarrow \bb J_{1}{}_\bB{}^\bA$.

The true power of the $R\mathcal T\mathcal T$-formulation, as employed in \cite{Beisert:2014hya} (see also \cite{Arutyunov:2006yd}), lies in the fact that the very $R$-matrix is a representation of the maps $\mathcal T$ defined above. Thus, in order to find either the Yangian (anti-)commutation relations or a representation of the $\bb T_\bB{}^\bA$s, one simply computes the Laurent series for $R=R(u,v)$. Specifically, denoting by $\pi_v$  a spectral-parameter-dependent representation of $\mathcal Y (\alg{gl}(p|q))$ onto $\text{End}(V)$ with $v\in\mathbbm{C}$ and setting 
\begin{subequations}
\begin{equation}
  T_\bB{}^\bA(u,v)\ :=\ \pi_v(\bb T_\bB{}^\bA(u))\quad\Longleftrightarrow\quad T_{m}{}_\bB{}^\bA(v)\ :=\ \pi_v(\bb T_{m}{}_\bB{}^\bA)
\end{equation}
we have
\begin{equation}\label{eq:repRfromT}
R(u,v)\ =\ (\mathbbmss{1}\otimes\pi_v){\mathcal T}(u)\ =\ E_\bA{}^\bB\otimes \sum_{m\geq 0} u^{-m}\,T_{m-1}{}_\bB{}^\bA(v)~. 
\end{equation}
\end{subequations}

\subsection{Algebraic formulation and representations}

Let us now move on and specialise to the Yangian $\mathcal{Y}(\mathfrak{gl}(1|1)_l)$. We shall first analyse the algebraic formulation of the $\bb T_\bB{}^\bA$s and then discuss explicit representations.

\paragraph{Algebraic formalism.}
 Let us first focus on the left representation $\pi_{\smallL,u_{\smallL}}$, and, consequently, on the left-left $R$-matrix $R_{\smallL\smallL}$ as given in \eqref{eq:RLL}. In order to be able to write (\ref{eq:repRfromT}) and derive abstract commutation relations of the generators $ \bb T_{m}{}_\bB{}^\bA$, we need to set (see \eqref{eq:defofex})
\begin{equation}
\begin{aligned}
u_{\smallL}\ :=\ \di\frac{h}{2}x^+_1\quad &\Rightarrow\quad  \di\frac{h}{2}x^-_1\ =\ u_{\smallL}+s -\frac{h^2s}{4u_{\smallL}^2}+\mathcal{O}(u_{\smallL}^{-3})~,\\
v_{\smallL}\ :=\ \di\frac{h}{2}x^+_2\quad &\Rightarrow\quad  \di\frac{h}{2}x^-_2\ =\ v_{\smallL}+s -\frac{h^2s}{4v_{\smallL}^2}+\mathcal{O}(v_{\smallL}^{-3})  ~
\end{aligned}
\end{equation}
and expand $R_{\smallL\smallL}$ for large $u_{\smallL}$. One finds that 
\begin{equation}
 \bb T_{-1}{}_\bB{}^\bA\ =\ \delta_\bB{}^\bA\bb U^{|\bB|}~.
\end{equation}
The $R\mathcal T\mathcal T$-relations \eqref{eq:RTTRel} involving $ \bb T_{-1}{}_\bB{}^\bA$ show it is central. This, in turn, implies the centrality of $\bb U$. Furthermore, the $R\mathcal T\mathcal T$-relations involving  $ \bb T_{0}{}_\bB{}^\bA$ yield the (anti-)commutation relations of a certain deformation of the Lie superalgebra $\mathfrak{gl}(1|1)_l$. In fact, to obtain the \linebreak(anti-)commutation relations of $\mathfrak{gl}(1|1)_l$,  one should re-define the generators and work with 
\begin{equation}\label{eq:DefOfJ0viaT0}
 \bb J_{0}{}_\bB{}^\bA\ :=\ \bb U^{-\frac12(|\bA|+|\bB|)}\bb T_{0}{}_\bB{}^\bA
 \end{equation}
  instead of $\bb T_{0}{}_\bB{}^\bA$.  Moreover, from level 1 upwards, the $R\mathcal T\mathcal T$-relations \eqref{eq:RTTRel} neither yield directly the (anti-)commutation relations of  $\mathfrak{gl}(1|1)_l$ nor those of its Yangian $\mathcal{Y}(\mathfrak{gl}(1|1)_l)$. For instance, one may check that 
\begin{equation}
\ccomm{\bb T_{1}{}_2{}^1,\bb T_{0}{}_1{}^2}\ =\ s\big(\!-\bb U\,\bb T_{1}{}_1{}^1+\bb T_{1}{}_2{}^2+\tfrac{1}{4}\bb T_{0}{}_2{}^1\bb T_{0}{}_1{}^2-\tfrac{1}{4}\bb T_{0}{}_1{}^2\bb T_{0}{}_2{}^1\big)~.
\end{equation}
Hence, to bring the (anti-)commutation relations into Yangian form we shall follow \cite{Beisert:2014hya} and, in addition to \eqref{eq:DefOfJ0viaT0}, consider
\begin{equation}\label{eq:DefOfJ1viaT0T1}
\bb J_{1}{}_\bB{}^\bA\ :=\ \bb U^{-\frac12(|\bA|+|\bB|)}\bb T_{1}{}_\bB{}^\bA-\tfrac12(-)^{(|\bA|+|\bC|)(\bB|+|\bC|)}\bb U^{-\frac12(|\bA|+|\bB|+2|\bC|)}\bb T_{0}{}_\bB{}^\bC \bb T_{0}{}_\bC{}^\bA~.
\end{equation}
The $\bb J_{m}{}_\bB{}^\bA$ we defined are slightly different from those reported in \cite{Beisert:2014hya} as we want the corresponding coproducts to be in the most symmetric frame.
 
Since the symmetry of the  $R$-matrix for $\AdS_3\times S^3\times M^4$  is a deformation of the Yangian $\mathcal Y(\mathfrak{gl}(1|1)_l\times\mathfrak{gl}(1|1)_r)$, we expect identifications that resemble the $\AdS_5\times S^5$ ones. Indeed, a short calculation shows that the combinations\fn{ The discussion for higher-level generators can be performed similarly.}
\begin{equation}\label{eq: generatorsinjojoamanifashion}
\begin{gathered}
\bb B_0\ :=\ \tfrac{2}{s}({\bb J}_0{}_1{}^1+{\bb J}_0{}_2{}^2)~,\qquad
\bb B_1\ :=\ \tfrac1s\big({\bb J}_1{}_1{}^1+{\bb J}_1{}_2{}^2)+{\bb J}_0{}_1{}^1+\tfrac12\big(\bb Q_0\bb S_0-\bb S_0\bb Q_0)+\tfrac 12\bb B_0\bb B_0~,\\
\bb H_0\ :=\ -{\bb J}_0{}_1{}^1+{\bb J}_0{}_2{}^2~,\qquad
\bb H_1\ :=\ -{\bb J}_1{}_1{}^1+{\bb J}_1{}_2{}^2+\tfrac 12\bb H_0\bb H_0+\tfrac s2\bb H_0~,\\
\bb Q_0\ :=\ \tfrac{1}{\sqrt{s}}{\bb J}_0{}_1{}^2~,\qquad
\bb Q_1\ :=\ \tfrac{1}{\sqrt{s}}{\bb J}_1{}_1{}^2+\tfrac 12\bb Q_0\bb H_0+\tfrac s2\bb Q_0~,\\
\bb S_0\ :=\ -\tfrac{1}{\sqrt{s}}{\bb J}_0{}_2{}^1~,\qquad
\bb S_1\ :=\ -\tfrac{1}{\sqrt{s}}{\bb J}_1{}_2{}^1+\tfrac 12\bb S_0\bb H_0+\tfrac s2\bb S_0,~\end{gathered}
\end{equation}
for the left copy $\mathfrak{gl}(1|1)_l$ in terms of the $\bb J_0{}_\bB{}^\bA$ and $\bb J_1{}_\bB{}^\bA$ as given in \eqref{eq:DefOfJ0viaT0} and \eqref{eq:DefOfJ1viaT0T1} obey ($m,n=0,1$)
\begin{equation}
  \begin{gathered}
    \comm{\bb B_0}{\bb Q_n}\ =\ -2 \,\bb Q_n~, \quad
    \comm{\bb B_0}{\bb S_n}\ =\ 2 \, \bb S_n~, \quad
    \acomm{\bb Q_m}{\bb S_n}\ =\ -\bb H_{m+n}~, \\
    \comm{\bb B_m}{\bb B_n}\ =\ 0~, \quad \comm{\bb H_m}{\cdot}\ =\ 0~, \quad 
    \acomm{\bb Q_m}{\bb B_n}\ =\ 
    \acomm{\bb S_m}{\bb S_n}\ =\ 0~, \\
    \comm{\bb B_{m+1}}{\bb Q_n} - \comm{\bb B_m}{\bb Q_{n+1}} + \acomm{\bb B_m}{\bb Q_n}\ =\ 0~, \quad
    \comm{\bb B_{m+1}}{\bb S_n} - \comm{\bb B_m}{\bb S_{n+1}} - \acomm{\bb B_m}{\bb S_n}\ =\ 0~.
  \end{gathered}
\end{equation}
The (anti-)commutation relations coincide precisely with \eqref{eq:Lie}. Here, $\bb B_1$ corresponds the first secret generator embedded in the Yangian we have found in the first part of the paper: see \eqref{eq:LL1-YG} and \eqref{eq:RL1-YG}. 

We would like to emphasise that the above (anti-)commutation relations are inherited from the $R\mathcal T\mathcal T$-relations: consequently, they are truly universal. Indeed, the generators given in \eqref{leftrep} and \eqref{eq:LL1-YG} are simply representations (by means of $\pi_{\smallL,u_{\smallL}}$) of these abstract ones, as we will demonstrate shortly.

The coproducts for the generators  \eqref{eq: generatorsinjojoamanifashion} correctly match the expected ones (see e.g. \eqref{eq:random1} and \eqref{eq:random2}). To verify this, one needs to make use of the identity 
\be{
\Delta\p{\bb T{}_\bB{}^\bA(u)}\ =\  \bb T{}_\bB{}^\bC(u)\otimes \bb T{}_\bC{}^\bA(u)
} 
induced by the $R$-matrix fusion relations.

The coproduct and commutation rules for 
\begin{equation}
\ss\ :=\ \ -\tfrac1s\big({\bb J}_1{}_1{}^1+{\bb J}_1{}_2{}^2)-{\bb J}_0{}_1{}^1
\end{equation}
are the same as those for the second secret generator, however, the representations $\pi_{\smallL,u_{\smallL}}({\ss})$ and $\pi_{\smallL,u_{\smallL}}(\hat {\mathfrak b})$ do not coincide. On the other hand, 
\begin{equation}
\hat {\mathfrak b}_L\ :=\ -\tfrac1s\big({\bb J}_1{}_1{}^1+{\bb J}_1{}_2{}^2)-{\bb J}_0{}_1{}^1-\tfrac12\id
\end{equation}
gives the correct representation, but not the right coproduct. The key to resolving this issue is to consider equivalence classes of generators as in \eqref{mappa}. This enables us to reproduce all the results from Section \ref{sec:YSS}.

%%%%%%%%%%%%%%%%%%%%%%%%%%%%%%%%%%%%%%%%%%%%%%%%%%%%%%%%%%%

\paragraph{Representations.}
By projecting the second leg of the $R$-matrix we obtain a representation for the abstract generators we found in the previous section.  In particular, upon Laurent-expanding the left-left $R$-matrix \eqref{eq:RLL} in the spectral parameter $u_{\smallL}$ around infinity, we obtain immediately from \eqref{eq:repRfromT} the following expressions for $T_{m}{}_\bB{}^\bA(v_{\smallL})$ for $m=-1,0$:
\begin{subequations}
\begin{equation}\label{eq:Trep-1}
\begin{gathered}
 T_{\smallL}{}_{\,-1}{}_1{}^1(v_{\smallL})\ =\ \id~,\qquad
 T_{\smallL}{}_{\,-1}{}_2{}^2(v_{\smallL})\ =\ U(v_{\smallL})\id~,\\
  T_{\smallL}{}_{\,-1}{}_1{}^2(v_{\smallL})\ =\ 0\ =\  T_{\smallL}{}_{\,-1}{}_2{}^1(v_{\smallL})
  \end{gathered}
\end{equation}
and 
\begin{equation}\label{eq:Trep0}
\begin{gathered}
T_{\smallL}{}_{\,0}{}_1{}^1(v_{\smallL})\ =\tfrac h4\gamma^2(v_{\smallL})\id + \tfrac s4\big(E_1{}^1+E_2{}^2\big)~,\quad
T_{\smallL}{}_{\,0}{}_2{}^1(v_{\smallL})\ =\ \gamma(v_{\smallL})\sqrt{\tfrac{h}{2}\,s\,U(v_{\smallL})}\, E_2{}^1~,\\
T_{\smallL}{}_{\,0}{}_1{}^2(v_{\smallL})\ =\ \gamma(v_{\smallL})\sqrt{\tfrac{h}{2}\,s\,U(v_{\smallL})}\,E_1{}^2,\quad
T_{\smallL}{}_{\,0}{}_2{}^2(v_{\smallL})\ =U(v_{\smallL})\big[-\tfrac h4\gamma^2(v_{\smallL})\id + \tfrac s4\big(E_1{}^1+E_2{}^2\big)\big]
\end{gathered}
\end{equation}
with
\begin{equation}
 U(v_{\smallL})\ :=\ \sqrt{\frac{x^+_2}{x^-_2}}\eand
  \gamma(v_{\smallL})\ :=\ \sqrt{\di(x^-_2-x^+_2)}
\end{equation}
\end{subequations}
All higher level generators can be found in the same spirit. By using (\ref{eq: generatorsinjojoamanifashion}), this precisely reproduces the representation of the first part of the paper, with the level-1 Yangian generators indeed given in the evaluation representation with spectral parameter $v_{\smallL}$, that is, $\gen{J}_{\smallL}{}_{\,1}{}_\bA{}^\bB=v_{\smallL}\gen{J}_{\smallL}{}_{\,0}{}_\bA{}^\bB$. Here, $\gen{J}_{\smallL}{}_{\,0,1}{}_\bA{}^\bB$ denotes the left representation of the generators in \eqref{eq: generatorsinjojoamanifashion}.

\subsection{Central Extensions}\label{centro}

Finally, we would like to comment on central extensions. We can show that, if now we take the right-left $R$-matrix \eqref{eq:RRL} and proceed with the $R\mathcal T\mathcal T$-formalism in that case, we are actually capable of reproducing the right copy $\alg{gl}(1|1)_r$ of the algebra $\alg{gl}(1|1)_l\times\alg{gl}(1|1)_r$. This means that the Laurent expansion of the right-left $R$-matrix \eqref{eq:RRL} in $u_R$ around zero yields the appropriate $T_{\smallR}{}_{\,m}{}_\bB{}^\bA(v_{\smallL})$ generators for the right copy of the factor algebra.\footnote{We choose an expansion around $u_{\smallR}=0$ as this is the point at which the right-left $R$-matrix becomes almost the identity needed to eventually get the Yangian charges. Let us notice that either large $u_{\smallL}$ or small $u_{\smallR}$ correspond to large $u_{\AdS_5}$ as utilised in \cite{Beisert:2014hya}.} The explicit calculation confirms that, as anticipated, they behave as their left partner.\footnote{For the right copy $\mathfrak{gl}(1|1)_r$, we shall use the very same combinations \eqref{eq: generatorsinjojoamanifashion} apart from $\bb B_1$ and $\hat\beta$ which we re-define as $\bb B_1\to \bb B_1-\bb H_0+\id$ and $\hat\beta\to\hat\beta+\bb H_0$. This (central) shift is necessary to match the evaluation representation for the right copy.}  We therefore refrain from repeating the whole procedure.  Instead, we shall rather derive the central extensions of the algebra by means of the $R\mathcal T\mathcal T$-relations \eqref{eq:RTTRel} via the expansion of $R_{RL}$:
\begin{subequations}
\begin{equation}
\begin{gathered}
\ccomm{\bb Q_{l0},\bb Q_{r0}}\ =\ \bb{ P}_0~,\qquad 
\ccomm{\bb{ Q}_{l1},\bb Q_{r0}}\ =\ \bb{P}_{l1}~,\qquad 
\ccomm{\bb{ Q}_{l0},\bb{Q}_{r1}}\ =\ \bb{P}_{r1}~,\\
\ccomm{\bb S_{l0},\bb S_{r0}}\ =\ \bb{K}_0~,
\qquad \ccomm{\bb{S}_{l1},\bb S_{r0}}\ =\ \bb{ K}_{l1}~,\qquad 
\ccomm{\bb{ S}_{l0},\bb{ S}_{r1}}\ =\ \bb{K}_{r1}~,
\end{gathered}
\end{equation}
with 
\begin{equation}
\begin{gathered}
\bb P_0 \ =\ \bb K_0 \ =\ \di\tfrac{h}{2}\p{\bb U-\bb U^{-1}}\eand 
\bb P_{l,r\,1}\ =\ \bb K_{l,r\,1}\ =\  \di\tfrac{h}{2}\bb U\,\bb H_{l,r\,0}~.
\end{gathered}
\end{equation}
\end{subequations}

In summary, the $R\mathcal T\mathcal T$-formulation does not only gives back the whole deformed Yangian but also puts left and right algebras on the same footing. 

Finally, we wish to emphasise that one should be capable of deriving these relations from the universal $R$-matrix $\cal{R}$ of the full algebra by means of
\begin{equation}
\mathcal{T}(u) \ = \ E_\bA{}^\bB\otimes \bb T_{\bB}{}^\bA(u) \ =\ E_\bA{}^\bB\otimes \sum_{m\geq0} u^{-m}  \bb T_{m-1}{}_{\bB}{}^\bA(u) \ = \ (\pi_u \otimes \mathbbmss{1}) \, {\cal{R}}~. 
\end{equation}
Here, $\pi_u$ denotes the suitable spectral-parameter-dependent representation onto $\text{End}(V)$. Turning the argument around, our treatment can give important insight into the issue of formulating a universal $R$-matrix for the current problem.

\section*{Acknowledgements}

We thank A.~Camobreco, T.~\L ukowski, and B.~Stefanski for fruitful discussions, and also the referee for useful comments. AP is supported in parts by the EPSRC under EP/K503186/1. AT thanks the EPSRC for funding under the First Grant EP/K014412/1 {\it Exotic quantum groups, Lie superalgebras and integrable systems}.

\bibliographystyle{latexeu}
%\bibliography{references}

\end{document}